\documentclass{svjour3}

\usepackage{cite}

\begin{document}

\title{Algebraic treatment of a simple model for the electromagnetic self-force}
\author{Francisco M. Fern\'{a}ndez}

\institute{INIFTA (UNLP,CCT La Plata-CONICET), Blvd. 113 y 64 S/N,
Sucursal 4, Casilla de Correo 16, 1900 La Plata, Argentina
\email{fernande@quimica.unlp.edu.ar}}

\date{Received: date / Accepted: date}

\maketitle

\begin{abstract}
The problem of the electromagnetic self-force can be studied in terms of a
quadratic PT-symmetric Hamiltonian. Here, we apply a straightforward
algebraic method to determine the regions of model-parameter space where the
quantum-mechanical operator exhibits real spectrum. An alternative point of
view consists of finding the values of the model parameters so that a
symmetric operator supports bound states.


\end{abstract}

\section{Introduction}

\label{sec:intro}

In a recent paper Bender and Gianfreda\cite{BG15} discussed a model for the
electromagnetic self-force on an oscillating charge particle. They showed
that the problems reported by Englert\cite{E80} were due to the fact that
the PT-symmetric Hamiltonian that describes the system exhibits broken PT
symmetry. The addition of two interacting terms to that Hamiltonian enabled
Bender and Gianfreda to analyse the regions of broken and unbroken PT
symmetry. To this end the authors resorted to the approach of Rossignoli and
Kowalski\cite{RK05} in order to rewrite the quadratic Hamiltonian in
diagonal form. The authors mentioned the possibility of using an alternative
approach for the same purpose\cite{F14} which is known since long ago\cite
{FC96}.

The purpose of this paper is to stress the fact that the application of the
latter algebraic method to quadratic Hamiltonians is extremely simple and
straightforward. In section~\ref{sec:model} we derive the regular or adjoint
matrix representation of the modified Hamiltonian operator for the
electromagnetic self-force\cite{BG15} and obtain the spectral frequencies in
terms of the model parameters. These frequencies, which are roots of the
characteristic polynomial for the matrix already mentioned, are suitable for
determining the regions in parameter space where the PT symmetry is broken.
Finally, in section~\ref{sec:conclusions} we summarize the main results and
draw conclusions.

\section{The quantum-mechanical model}

\label{sec:model}

From the pair of classical equations of motion proposed by Englert\cite{E80}%
, Bender and Gianfreda\cite{BG15} derived the Hamiltonian function
\begin{equation}
H_{c}=\frac{p_{x}p_{w}-p_{y}p_{z}}{m\tau }+\frac{2p_{z}p_{w}}{m\tau ^{2}}+%
\frac{wp_{y}+zp_{x}}{2}-\frac{mzw}{2}+kxy,  \label{eq:H_c}
\end{equation}
where $q=x,y,z,w$ are suitable coordinates and $p_{q}$ the corresponding
conjugate momenta. The quantum-mechanical version of this operator is
PT-symmetric but its eigenvalues are not real because the PT symmetry is
broken for all $m,\tau ,k$. To overcome this difficulty Bender and Gianfreda
added two terms and obtained the modified Hamiltonian operator
\begin{eqnarray}
H &=&\frac{B\left( wp_{z}-zp_{w}\right) }{m\tau }+\frac{2p_{z}p_{w}}{m\tau
^{2}}+\frac{p_{x}p_{w}-p_{y}p_{z}}{m\tau }-\frac{mzw}{2}+\frac{wp_{y}+zp_{x}%
}{2}+kxy  \nonumber \\
&&+\frac{A\left( x^{2}+y^{2}\right) }{2},  \label{eq:H}
\end{eqnarray}
where the coordinates and momenta satisfy the quantum-mechanical commutation
relations $[q,p_{q}]=i$. In this way those authors were able to show that
the resulting PT-symmetric Hamiltonian exhibits real spectrum for suitable
values of $A$ and $B$. In particular, the PT symmetry is broken for the case
$A=B=0$ that leads to $H_{c}$.

The simplest way of determining the regions of broken PT symmetry is to
convert $H$ into a suitable diagonal form. Bender and Gianfreda resorted to
the approach proposed by Rosignoli and Kowalski\cite{RK05} and in what
follows we apply a well known algebraic approach[F96] proposed recently by
Fern\'{a}ndez\cite{F14} for the treatment of a simple model for optical
resonators\cite{BGOPY13}.

The algebraic approach is based on the construction of the adjoint or
regular matrix representation of $H$ in the operator basis $\left\{
O_{1},O_{2},\ldots ,O_{8}\right\} =\left\{
x,y,z,w,p_{x},p_{y},p_{z},p_{w}\right\} $. The matrix elements $H_{ij}$ are
given by the coefficients of the commutator relations
\begin{equation}
\lbrack H,O_{i}]=\sum_{j=1}^{8}H_{ji}O_{j},\;i=1,2,\ldots ,8.
\label{eq:[H,O_i]}
\end{equation}
A straightforward calculation leads to
\begin{equation}
\mathbf{H}=\left(
\begin{array}{llllllll}
0 & 0 & 0 & 0 & iA & ik & 0 & 0 \\
0 & 0 & 0 & 0 & ik & iA & 0 & 0 \\
-\frac{i}{2} & 0 & 0 & \frac{iB}{m\tau } & 0 & 0 & 0 & -\frac{im}{2} \\
0 & -\frac{i}{2} & -\frac{iB}{m\tau } & 0 & 0 & 0 & -\frac{im}{2} & 0 \\
0 & 0 & 0 & -\frac{i}{m\tau } & 0 & 0 & \frac{i}{2} & 0 \\
0 & 0 & \frac{i}{m\tau } & 0 & 0 & 0 & 0 & \frac{i}{2} \\
0 & \frac{i}{m\tau } & 0 & -\frac{2i}{m\tau ^{2}} & 0 & 0 & 0 & \frac{iB}{%
m\tau } \\
-\frac{i}{m\tau } & 0 & -\frac{2i}{m\tau ^{2}} & 0 & 0 & 0 & -\frac{iB}{%
m\tau } & 0
\end{array}
\right) .  \label{eq:H_mat}
\end{equation}
The spectral frequencies of the problem are the eigenvalues of this matrix.
The secular equation $|\mathbf{H}-\lambda \mathbf{I}|=0$ (where $\mathbf{I}$
is the $8\times 8$ identity matrix) yields the characteristic polynomial

\begin{equation}
\left( m^{2}\tau ^{2}\xi -B^{2}+m^{2}\right) \left( m^{2}\tau ^{2}\xi
^{3}+\xi ^{2}\left( m^{2}-B^{2}\right) +\xi \left( 2AB-2km\right)
-A^{2}+k^{2}\right) =0,  \label{eq:charpoly_xi}
\end{equation}
where $\xi =\lambda ^{2}$. One of the roots is

\begin{equation}
\xi =\frac{B^{2}-m^{2}}{m^{2}\tau ^{2}},  \label{eq:xi_1}
\end{equation}
and the remaining three ones are solutions to the cubic equation

\begin{equation}
m^{2}\tau ^{2}\xi ^{3}+\left( m^{2}-B^{2}\right) \xi ^{2}+2\left(
AB-km\right) \xi -A^{2}+k^{2}=0.  \label{eq:charpoly_xi_2}
\end{equation}
When the spectrum is real the frequencies are real and the four roots $\xi
_{j}$ are positive. Thus the region in parameter space where the spectrum is
real is determined by the set of equations $\left\{
B^{2}>m^{2},\;A^{2}>k^{2},\;AB>km\right\} $.

Bender and Gianfreda\cite{BG15} obtained exactly the same equations (\ref
{eq:xi_1}) and (\ref{eq:charpoly_xi_2}) by means of an alternative approach
based on the creation and annihilation operators\cite{RK05}. In our opinion
the adjoint or regular matrix representation of the Hamiltonian operator
provides a more straightforward route. In addition to it, the algebraic
approach is somewhat more general because it applies to any Hamiltonian that
satisfies the commutator relations (\ref{eq:[H,O_i]})\cite{FC96}.

\section{Conclusions}

\label{sec:conclusions}

Any Hamiltonian that is a quadratic function of the coordinates and momenta
or of the creation and annihilation operators is exactly solvable. One way
of obtaining its spectrum is to convert it into a diagonal form by means of
a linear combination of the relevant dynamical variables\cite{RK05}. The
algebraic method\cite{F14,FC96} is a particularly simple approach because
the construction of the adjoint or regular matrix representation is
straightforward. It only requires the calculation of simple commutation
relations like (\ref{eq:[H,O_i]}).

It is also worth mentioning that the quantum-mechanical versions of $H_{c}$
and $H$ are symmetric operators. They satisfy $\left\langle f\right| H\left|
g\right\rangle =\left\langle g\right| H\left| f\right\rangle ^{*}$ for any
pair of square integrable functions $f(\mathbf{q})$ and $g(\mathbf{q})$. If $%
\psi $ is a normalizable eigenfunction of $H$ ($H\psi =E\psi $, $%
\left\langle \psi \right| \left. \psi \right\rangle <\infty $) then $E$ is
real. Therefore, in the present case, looking for the regions in parameter
space of unbroken PT symmetry is equivalent to looking for the regions where
$H$ supports bound states (as Bender and Gianfreda\cite{BG15} did explicitly
for the ground state). The same situation emerged in the case of the optical
resonators mentioned above\cite{F14,BGOPY13}.


\begin{thebibliography}{9}
\bibitem{BG15}  Bender, C. and Gianfreda, M.: PT-symmetric interpretation of
the electromagnetic self-force. J. Phys. A 48, 34FT01 (2015).

\bibitem{E80}  Englert, B.-G.: Quantization of the radiation-damped harmonic
oscillator. Ann. Phys. 129, 1-21 (1980).

\bibitem{RK05}  Rossignoli, R. and Kowalski, A. M.: Complex modes in
unstable quadratic bosonic forms. Phys. Rev. A 72, 032101 (2005).

\bibitem{F14}  Fern\'{a}ndez, F. M.: Algebraic Treatment of PT -Symmetric
Coupled Oscillators. Int. J. Theor. Phys. DOI 10.1007/s10773-10014-12201-y
(2014). arXiv:1402.4473 [quant-ph].

\bibitem{FC96}  Fern\'{a}ndez, F. M. and Castro, E. A.: Algebraic Methods in
Quantum Chemistry and Physics, Mathematical Chemistry Series, CRC, Boca
Raton, New York, London, Tokyo (1996).

\bibitem{BGOPY13}  Bender, C. M., Gianfreda, M., \"{O}zdemir, S. K., Peng,
B., and Yang, L.: Twofold transition in PT-symmetric coupled oscillators.
Phys. Rev. A 88, 062111 (2013).
\end{thebibliography}
\end{document}